\documentclass[pra,aps,twocolumn,nopacs,superscriptaddress,nofootinbib]{revtex4}
\usepackage[T1]{fontenc}
\usepackage[latin1]{inputenc}
\usepackage{lmodern}
\usepackage{graphicx}
\usepackage{dcolumn}
\usepackage{bm}
\usepackage{amsmath}
\usepackage{amssymb}
\usepackage{pst-all}
\usepackage{psfrag}
\usepackage{epsfig}

\def\ii{{\rm i}}  \def\ee{{\rm e}}  
      
    \def\Eb{{\bf E}}  \def\rb{{\bf r}}
\def\xx{\hat{\bf x}}    
 
    \def\EF{{E_{\rm F}}}
\def\vF{{v_{\rm F}}}   \def\EFT{{E_{\rm F}^T}}
\begin{document}

\title{Resonant Visible Light Modulation with Graphene}
\author{Renwen~Yu}
\affiliation{ICFO - Institut de Ciencies Fotoniques, Mediterranean Technology Park, 08860 Castelldefels (Barcelona), Spain}
\author{Valerio~Pruneri}
\affiliation{ICFO - Institut de Ciencies Fotoniques, Mediterranean Technology Park, 08860 Castelldefels (Barcelona), Spain}
\affiliation{ICREA - Instituci\'o Catalana de Recerca i Estudis Avan\c{c}ats, Barcelona, Spain}
\author{F.~Javier~Garc\'{\i}a~de~Abajo}
\affiliation{ICFO - Institut de Ciencies Fotoniques, Mediterranean Technology Park, 08860 Castelldefels (Barcelona), Spain}
\affiliation{ICREA - Instituci\'o Catalana de Recerca i Estudis Avan\c{c}ats, Barcelona, Spain}
\email{javier.garciadeabajo@icfo.es}

\begin{abstract}
\textbf{Fast modulation and switching of light at visible and near-infrared (vis-NIR) frequencies is of utmost importance for optical signal processing and sensing technologies. No fundamental limit appears to prevent us from designing wavelength-sized devices capable of controlling the light phase and intensity at gigaherts (and even terahertz) speeds in those spectral ranges. However, this problem remains largely unsolved, despite recent advances in the use of quantum wells and phase-change materials for that purpose. Here, we explore an alternative solution based upon the remarkable electro-optical properties of graphene. In particular, we predict unity-order changes in the transmission and absorption of vis-NIR light produced upon electrical doping of graphene sheets coupled to realistically engineered optical cavities. The light intensity is enhanced at the graphene plane, and so is its absorption, which can be switched and modulated via Pauli blocking through varying the level of doping. Specifically, we explore dielectric planar cavities operating under either tunneling or Fabry-Perot resonant transmission conditions, as well as Mie modes in silicon nanospheres and lattice resonances in metal particle arrays. Our simulations reveal absolute variations in transmission exceeding $90\%$ as well as an extinction ratio $>15\,$dB with small insertion losses using feasible material parameters, thus supporting the application of graphene in fast electro-optics at vis-NIR frequencies.}
\end{abstract}
\date{\today}
\maketitle


\section{INTRODUCTION}

\begin{figure*}
\begin{center}
\includegraphics[width=120mm,angle=0,clip]{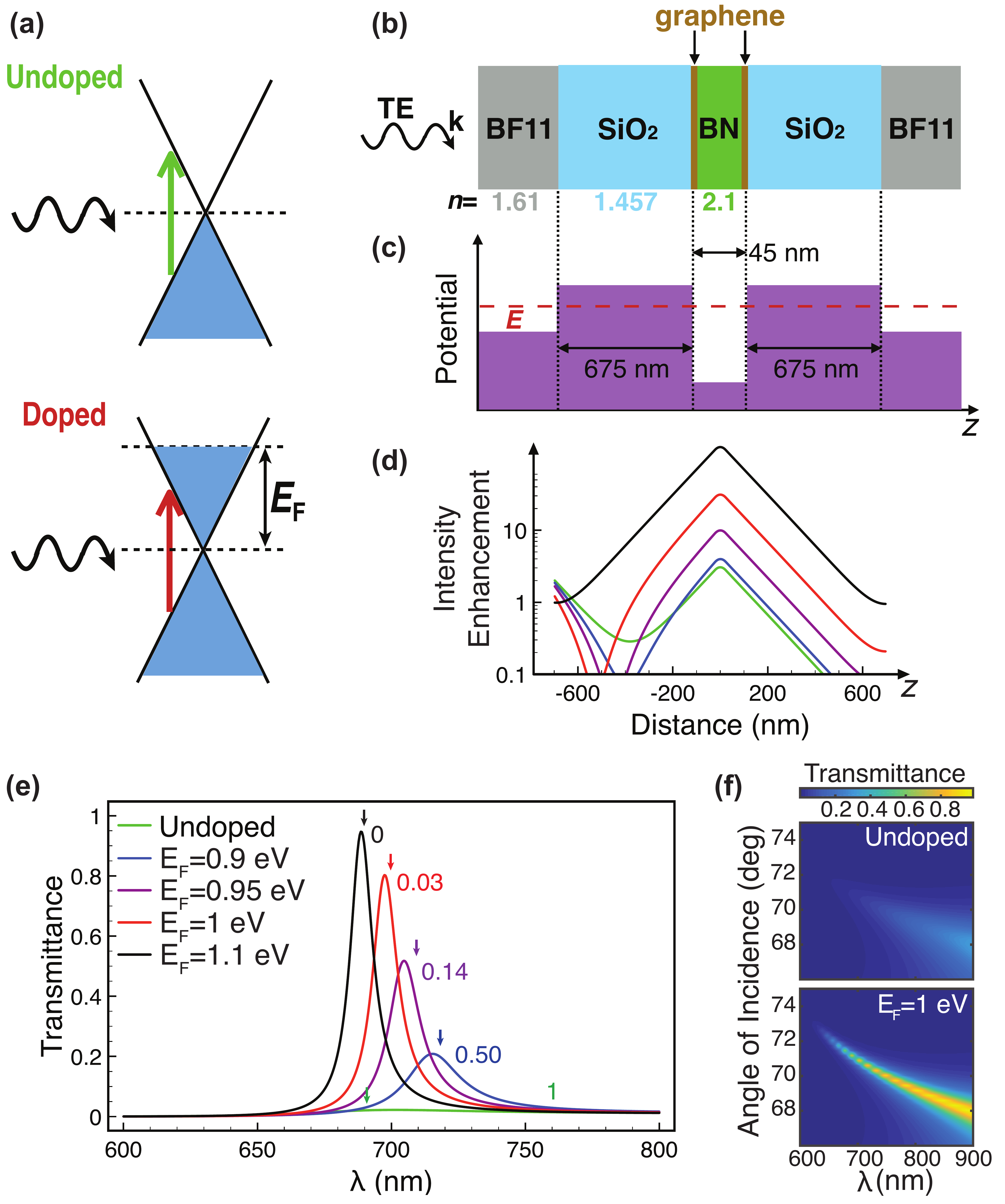}
\caption{{\bf Graphene optical switch based on resonant tunneling transmission.}
{\bf (a)} Doping-induced absorption switching effect used in this work: we compare undoped graphene (upper scheme, Fermi level at the Dirac point), which can absorb photons (vertical arrow) over a broad spectral range via interband electron transitions, and doped graphene (lower scheme), in which Pauli exclusion blocks photon absorption when the Fermi energy $E_{\rm F}$ exceeds half the photon energy. {\bf (b)} Planar multilayer structure considered for resonant tunneling transmission of light, including a central BN planar waveguide (not to scale) and two single-layer graphene films intercalated at the BN/SiO$_2$ interfaces. {\bf (c)} Potential in the equivalent Schr\"odinger model (see main text and Methods). {\bf (d)} Electric field intensity normalized to the external light intensity for an incidence angle of $71^\circ$ and a free-space wavelength of 689\,nm. Light is $s$ (TE) polarized and incident from the left. Results for different levels of doping are offered (see legend in (e)). {\bf (e)} Transmission spectra of the multilayer structure at $71^\circ$ incidence for different levels of doping. The transmission maxima are in excellent agreement with the analytical expression offered in the Methods section (see arrows). The numerical labels correspond to the ratio ${\rm Re}\{\sigma\}/(e^2/4\hbar)$ evaluated at a wavelength of 689\,nm. {\bf (f)} Transmission as a function of incidence angle and wavelength for doped and undoped graphene.} \label{Fig1}
\end{center}
\end{figure*}

Graphene --the two-dimensional (2D) honeycomb lattice of carbon atoms-- exhibits extraordinary optoelectronic properties derived from its peculiar band structure of massless charge carriers \cite{CGP09}. Notably, its optical absorption can be switched on/off via electrical doping. In its undoped state it absorbs a fraction $\pi\alpha\approx2.3\%$ of the incident light \cite{NBG08,MSW08} over a broad spectral range within the vis-NIR as a result of direct electron-hole pair transitions between its lower occupied Dirac cones and the upper unoccupied cones (two inequivalent ones in every Brillouin zone \cite{W1947,CGP09}). In contrast, when electrically doped, an optical gap is opened that suppresses vertical optical transitions for photon energies below $2|\EF|$, where $\EF$ is the change in Fermi energy relative to the undoped state (see Fig.\ \ref{Fig1}a). In practice, values of $\EF$ as high as 1\,eV can be obtained through electrical gating \cite{CPB11}, therefore enabling the modulation of light absorption down to the visible regime. Chemical methods permit achieving even higher levels of doping \cite{KWB12}, which could be combined with additional electrostatically induced variations of $\EF$ around a high bias point to reach control over shorter light wavelengths.

Fast light modulation at vis-NIR frequencies can find application in optical signal processing and interconnect switching, where there is a great demand for integrated wavelength-sized devices capable of operating at terahertz commutation rates. The extraordinary electro-optical response of graphene provides a key ingredient for the realization of these types of devices. However, the exploitation of atomically thin carbon films for light modulation faces the problem of their relatively weak interaction with light. A possible solution to enhance this interaction is to use the intrinsic plasmons that show up in the optical gap of this material when it is highly doped \cite{VE11,paper176,NGG11,JGH11,FAB11,BPV12,paper196,FRA12,paper212,BJS13,paper230,JBS14,paper235}. 
Resonant coupling to graphene plasmons can even result in complete optical absorption \cite{paper182}, as exemplified by the observation of large tunable light modulation at mid-IR frequencies in periodically nanostructured graphene \cite{paper230,JBS14}. The extension of this strategy down to the vis-NIR spectral domain remains a challenge, as it requires to laterally pattern the carbon film with $<10\,$nm features, which are currently unattainable through conventional lithographies, although chemical self-assembly might offer a viable way of producing the required structures \cite{M14}.

An alternative solution consists in amplifying the absorption of undoped graphene either by increasing the region over which light interacts with it or by coupling the carbon film to an optical cavity of high quality factor (i.e., by trapping light during long times near the graphene). A broadband modulator has been demonstrated with the former approach by exposing a long path of an optical waveguide to electrically gated graphene \cite{LYU11}. Additionally, coupling to photonic cavities has been explored using plasmonic structures, photonic-crystals, and metamaterials \cite{ECN12,GMG12,YKG13,LY13,MKA13,paper231,ECK14}. For example, monolayer graphene integrated with metallic metasurfaces has been used to control the optical response (resonance position, depth, and linewidth) at mid-IR frequencies \cite{YKG13,LY13,MKA13,ECK14}. Similarly, large intensity modulations ($>30\%$) of mid-IR light over a $600\,$nm bandwidth have been reported in graphene-loaded plasmonic antennas \citep{YKG13}. Additionally, a resonance wavelength shift $\sim2\,$nm accompanied by a 4-fold variation in reflectivity has been observed in the NIR by coupling graphene to a photonic crystal cavity \cite{GMG12,MKV13}. Enhanced visible light absorption in graphene has also been demonstrated (without modulation) by combining monolayer graphene with metamaterials \cite{PLS10}, gold nanovoid arrays \cite{ZSS13}, and photonic waveguides \cite{PF14}, as well as by coupling multilayer graphene under total internal reflection \cite{PML13}.

In this work, we study four different mechanisms that produce resonant enhancement in the absorption of undoped graphene over the vis-NIR spectral domain, thus serving as optical modulators with large depth in that frequency range. Specifically, we focus on the coupling of graphene to (1) resonant tunneling transmission cavities, (2) resonant Fabry-Perot cavities, (3) Mie modes of silicon spheres, and (4) lattice resonances in periodic particle arrays, which we investigate by calculating reflection, transmission, and absorption spectra of structures containing either doped or undoped graphene films. We predict modulation depths in vis-NIR light transmission exceeding $90\%$, with small insertion losses, thus revealing the potential of graphene for fast electro-optics within such a technologically important range of optical frequencies.

\section{RESULTS AND DISCUSSION}

\subsection{Graphene optical switch based upon resonant tunneling transmission}

We illustrate the concept of resonant switching and modulation of graphene absorption by coupling to a high-quality-factor planar cavity. In particular, we consider the multilayer structure depicted in Fig.\ \ref{Fig1}b, consisting of a high-refractive-index BN planar waveguide ($n_{\rm BN}=2.1$) flanked by low-index silica spacers ($n_{{\rm SiO}_2}=1.457$). The waveguide hosts guided modes that can be resonantly coupled to light of well-defined parallel wave vector (i.e., for a collimated incident beam). In our case, light is incident from the left under total internal reflection conditions at the BF11-SiO$_2$ interface ($n_{\rm BF11}=1.61$). The evanescent spill out of light intensity penetrating inside the left silica spacer can reach the BN waveguide, where it is amplified to further extend towards the rightmost interface. In the absence of absorption, full transmission can always be achieved at a resonant wavelength that depends on incidence angle. This phenomenon, known as resonant tunneling transmission, was previously explored with electron waves \cite{CET1974}. There is complete analogy between TE light propagation in the planar structure under consideration and the evolution of an electron according to Schr\"odinger equation \cite{paper161} (see Methods). The equivalent electron has energy $E$ and evolves along a potential profile as shown in Fig.\ \ref{Fig1}c. The latter is directly related to the refractive index, with higher index corresponding to lower values of the potential. The presence of a bound state is always guaranteed in a 1D cavity, and so is the existence of a full transmission resonance when this bound state lies inside the potential barrier \cite{CET1974}. Under complete-transmission conditions, the intensity has to decay exponentially from the waveguide to the far medium (i.e., across the rightmost silica barrier), to reach the same value as the incident intensity, so that the near field has to be strongly amplified at the central waveguide. This type of enhancement, which is clearly illustrated in Fig.\ \ref{Fig1}d, has been experimentally corroborated by measuring a $>100$-fold increase in the fluorescence from quantum dots placed near the central waveguide under resonance conditions in a structure similar to the one considered here \cite{paper161}. This effect can find application to sensing, for example by replacing the waveguide by a high-index fluid with dispersed analytes in it. Instead, in this article we use the resonant tunneling transmission concept to amplify the effect of absorption taking place at the graphene.

The structure under consideration (Fig.\ \ref{Fig1}b) contains a graphene film on either side of the central BN waveguide. Besides its high index of refraction, the choice of BN for the central waveguide is convenient because this combination of materials is compatible with high-quality graphene \cite{DYM10}, which can be realistically described with the models for the conductivity $\sigma$ discussed in the Methods section. Nevertheless, we assume a conservative value of the graphene mobility throughout this work, $\mu=2000\,$cm$^2/($V\,s$)$. When the carbon layer is highly doped ($\EF=1.1\,$eV), it becomes nearly lossless (i.e., small ${\rm Re}\{\sigma\}$) at the waveguide resonance wavelength, so that the peak transmission reaches $\sim95\%$ (Fig.\ \ref{Fig1}e) and the light intensity enhancement at the waveguide exceeds a factor of 140. In contrast, in the undoped state, the carbon layer becomes lossy (i.e., nearly real $\sigma\approx e^2/4\hbar$), so the enhancement is strongly suppressed, and the transmission drops to very small values. The extinction ratio (i.e., the ratio of transmissions in doped and undoped states) is $>15\,$dB. The transmission can be in fact tuned continuously between these two extreme values by varying the level of doping (see Fig.\ \ref{Fig1}e). The decrease in transmission produced when moving from highly doped to undoped graphene is due to both absorption and reflection, as the local change in the response of the carbon layer produces a departure from the conditions of resonant tunneling. Actually, reflection accounts for the bulk of the depletion in transmission, as shown in the supplementary Fig.\ \ref{FigS1}. This can be exploited to simplify the structure, which still undergoes unity-order modulation of the reflection upon graphene doping after removing the rightmost BF11 out-coupling medium (see supplementary Fig.\ \ref{FigS3}). Even more, only a single graphene layer is needed to modulate the structure (see supplementary Fig.\ \ref{FigS2}).

The wavelength of operation of this modulator is essentially determined by the waveguide mode, as coupling to the BF11 media is just producing a slight shift. Understandably, the reflection minimum is observed to be only mildly modified when the rightmost glass is removed (cf. supplementary Figs.\ \ref{FigS2} and \ref{FigS3}). Then, we find it useful to derive a simple analytical expression for the variation of the waveguide resonance wavelength (see Eq.\ (\ref{lambdares}) in the Methods section), in which the graphene conductivity enters to first order as $\propto{\rm Im}\{\sigma\}$. The role of the real and imaginary parts of $\sigma$ is thus clearly established: the former determines the depth of the modulation, whereas ${\rm Im}\{\sigma\}$ is responsible for the resonance shift. The resonance wavelengths given by Eq.\ (\ref{lambdares}) are indicated by downwards arrows in Fig.\ \ref{Fig1}e, in excellent agreement with the observed transmission maxima. Notice the initial redshift with increasing doping, followed by a blueshift back to the original position in the perfectly transmitting structure, which essentially mimics the evolution of ${\rm Im}\{\sigma\}$ with doping.

Obviously, the resonance wavelength also depends on the angle of incidence and it can be pushed down to the visible regime (Fig.\ \ref{Fig1}f), although the maximum transmission decreases towards smaller wavelengths due to the gradual involvement of interband transitions in the graphene.

In a realistic device, the two graphene layers of Fig.\ \ref{Fig1}b could be biased with a relative potential difference $V$, so that they will reach a Fermi energy $|\EF|=\hbar\vF\sqrt{V\epsilon_{\rm BN}/4d_{\rm BN}}$, where $\vF\approx10^6\,$m/s is the Fermi velocity in the carbon layer, while $\epsilon_{\rm BN}$ and $d_{\rm BN}$ are the static permittivity and thickness of the BN layer. For $d_{\rm BN}\sim45\,$nm, a value of $\EF=1\,$eV is obtained with potentials $\sim4\,$V. For the single graphene-layer structures noted above (see supplementary Figs. \ref{FigS2} and \ref{FigS3}), doping could be introduced through the addition of a transparent electrode.

\begin{figure}
\begin{center}
\includegraphics[width=80mm,angle=0,clip]{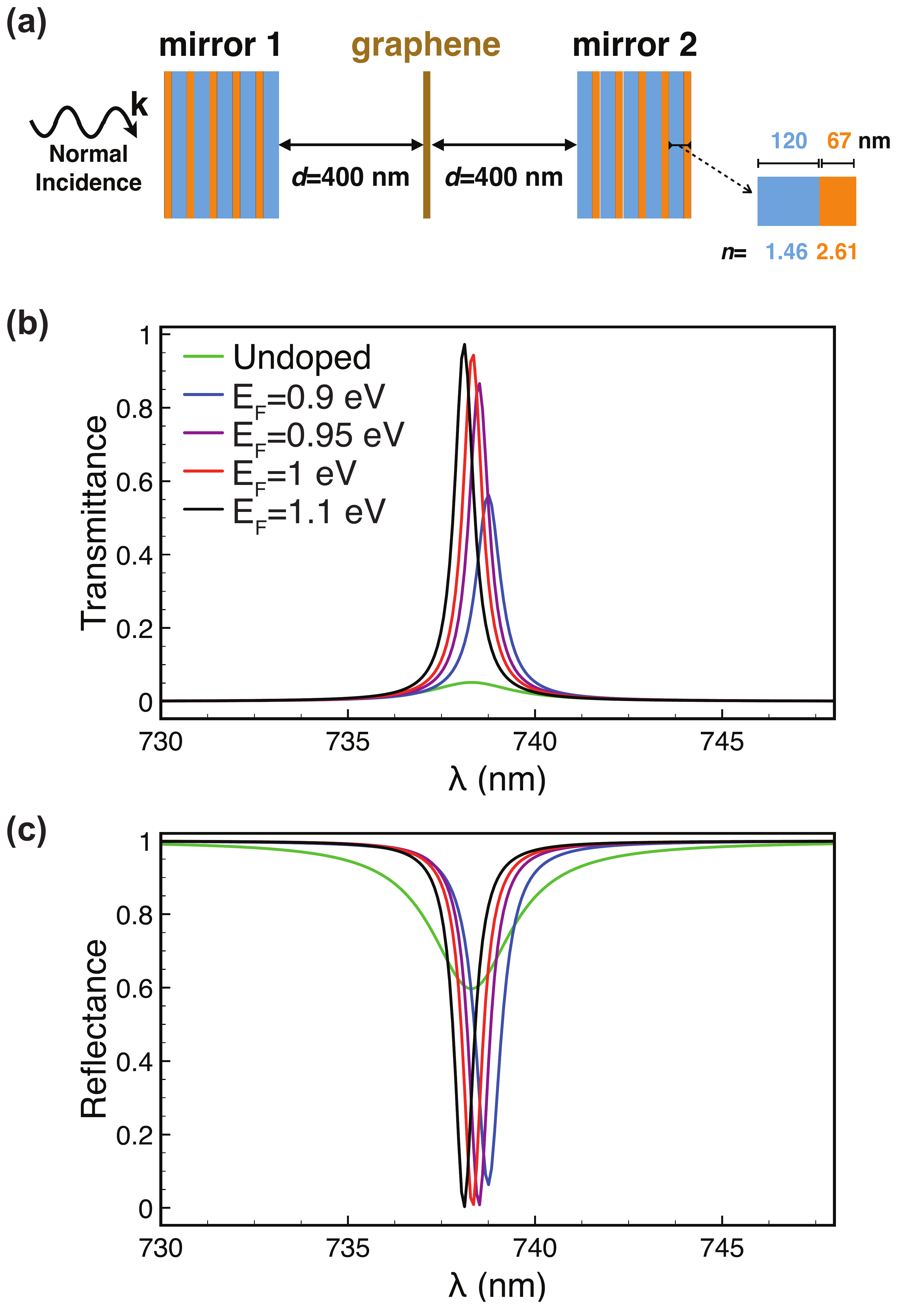}
\caption{{\bf Graphene optical switch based on resonant Fabry-Perot transmission.}
{\bf (a)} Fabry-Perot resonator incorporating a tunable graphene layer inside the cavity flanked by two Bragg mirrors (see inset for a sketch of the period and labels for geometrical and optical parameters). {\bf (b,c)} Normal-incidence transmittance (b) and reflectance (c) for different levels of doping. The cavity is filled with air, but similar performance is achieved with a glass-filled narrower cavity (see supplementary Fig.\ \ref{FigS4}).} \label{Fig2}
\end{center}
\end{figure}

\subsection{Graphene optical switch based upon a Fabry-Perot resonator}

The concept of the tunneling structure in Fig.\ \ref{Fig1} can be extrapolated to other types of resonators in which the incident field also undergoes a large enhancement at a position decorated with graphene. A particularly convenient implementation of this idea is presented in Fig.\ \ref{Fig2}, as it allows operating under normal incidence conditions. More precisely, we replace the tunneling structure by a Fabry-Perot (FP) frequency-selective filter, consisting of a cavity flanked by two non-absorbing, nearly perfectly reflecting mirrors. In practical devices, one generally uses Bragg mirrors such as those sketched in Fig.\ \ref{Fig2}a, which are easy to fabricate by multilayer deposition. We consider a separation between the FP mirrors that produces a single resonant transmission peak in the $730-748\,$nm spectral region. At resonance, light is trapped inside the cavity, so it makes many passes through it before escaping, thus generating a large field enhancement at several interference nodes. We place the graphene at one of those nodes. A similar strategy has been recently followed for all-optical nonlinear NIR light modulation \cite{FMC14}. An interplay between absorption (imaginary part of the susceptibility) and polarization (real part) in the graphene is then taking place, leading to large (but not totally complementary) modulations in reflection and transmission, similar to those discussed above for the tunneling device. Incidentally, similar performance is obtained by filling the cavity with glass and reducing its size (see supplementary Fig.\ \ref{FigS4}), thus configuring a more robust structure. We have also verified (results not shown) that further reduction of the cavity leads to a 1D photonic crystal that exhibits a normal-incidence gap, in which a localized optical mode exists due to the addition of an {\it impurity} (i.e., the cavity itself); the graphene can then couple to this localized mode to produce an even more compact light modulator. We find it interesting that the cavity is unaffected if the graphene is placed at an antinode of the interference standing wave inside the cavity (see supplementary Fig.\ \ref{FigS5}), as this opens the possibility of using an optically inactive graphene layer located at an antinode and serving as a gate with which to dope the other graphene layer placed at a node.

\begin{figure}
\begin{center}
\includegraphics[width=80mm,angle=0,clip]{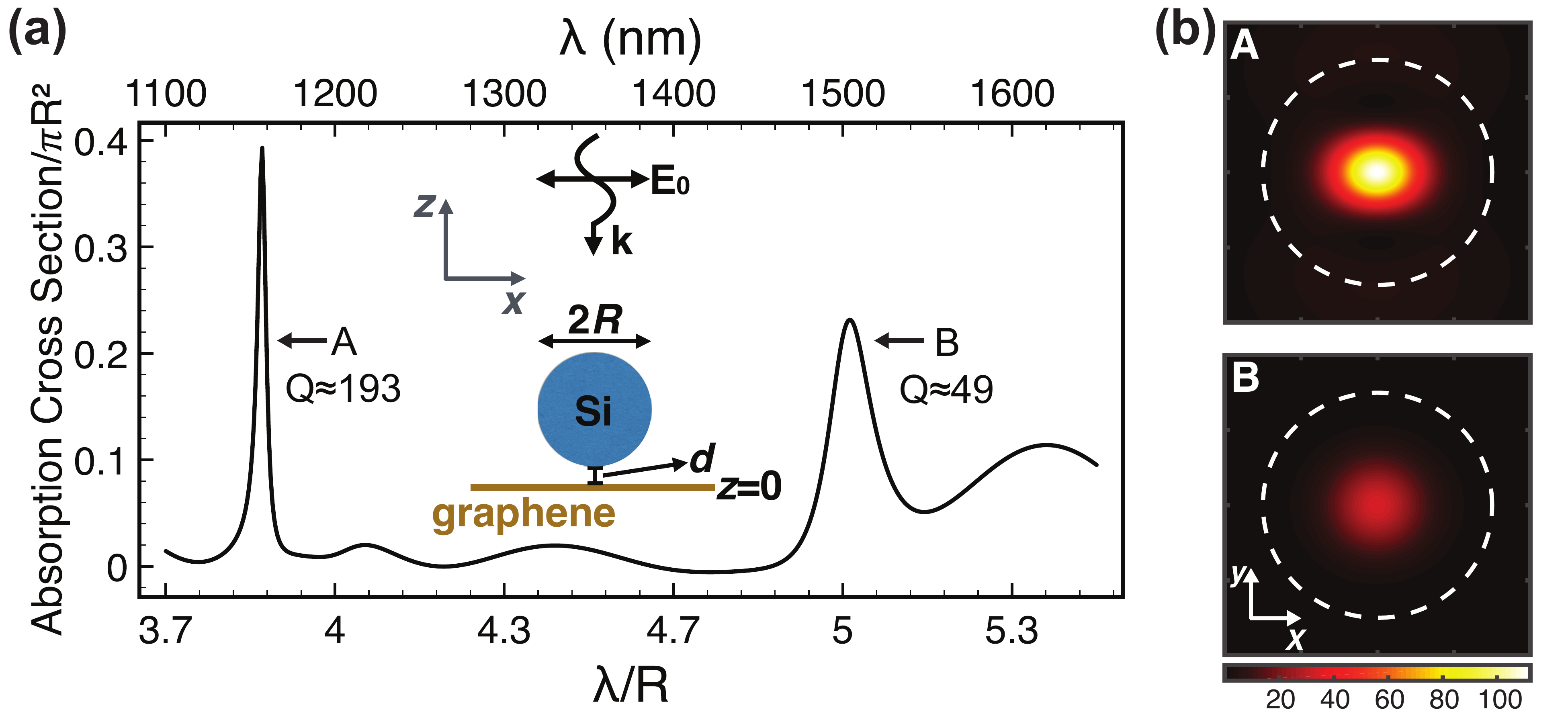}
\caption{{\bf Graphene absorption enhancement by coupling to Mie resonances.} {\bf (a)} Absorption cross section normalized to the projected sphere area ($\pi R^{2}$), estimated for the silicon-sphere/undoped-graphene system shown in the inset using Eq.\ (\ref{SAM}) and Mie theory. We plot the increase in absorption due to the presence of the sphere. The silicon/graphene separation is $d=R/150$. The upper scale corresponds to a sphere radius $R=300\,$nm and $d=2\,$nm. The incident electric field is along the $x$ direction. {\bf (b)} Parallel electric-field intensity enhancement $(|E_x|^2+|E_y|^2)/|E_0|^2$ at the graphene plane for the two Mie resonances labeled A and B in (a). The quality factors $Q$ of these resonances are also indicated in (a).} \label{Fig3}
\end{center}
\end{figure}

\subsection{Enhanced graphene optical absorption and switching by coupling to Mie cavities}

Figure\ \ref{Fig3}a represents the change in the absorption cross section undergone by a layer of undoped graphene when we place a silicon sphere ($\epsilon=12$) in its vicinity. These types of silicon colloids have been recently synthesized and used as excellent photonic cavities \cite{GFA14}. The increase in absorption cross section $\delta\sigma^{\rm abs}$ remains a small fraction of the extinction produced by the sphere in this configuration (e.g., 6.1\% and 2.7\% for the Mie modes labeled A and B in Fig.\ \ref{Fig3}a), so we approximate it as
\begin{equation}
\delta\sigma^{\rm abs}\approx\pi\alpha\int dxdy\;|\Eb_\parallel/E_0|^2,
\label{SAM}
\end{equation}
where $E_\parallel$ is the parallel component of the electric field scattered by the sphere alone, $E_0$ is the incident field, and we integrate over the graphene plane. The field $\Eb_\parallel$ is obtained from Mie theory \cite{M1908}. This approximate method yields similar results as the change in elastic (dark-field) scattering due to doping, calculated from a rigorous modal expansion for the sphere-graphene system (see supplementary Fig.\ \ref{FigS6}). In Fig.\ \ref{Fig3} the cross section is normalized to the projected area of the sphere $\pi R^2$ and the wavelength is normalized to the sphere radius $R$, so that this plot is independent of $R$, apart from the relatively small variations of the permittivity of silicon over the NIR. Despite the subwavelength size of the particle, its high $\epsilon$ allows it to trap light within Mie modes of high quality factor ($Q\approx193$ and $49$ in modes A and B, see Fig.\ \ref{Fig3}a), giving rise to large local enhancements of the near-field intensity at the plane of the graphene (see Fig.\ \ref{Fig3}b). This in turn boots the absorption, which takes remarkably large values, with a peak increase in cross section reaching $\sim40\%$ of the projected area of the sphere. Interestingly, the spatial distribution of absorption (proportional to the intensity plotted in Fig.\ \ref{Fig3}b) is strongly confined to the near-contact region, which could be exploited for engineering the spatial distribution of optically induced heat deposition, as well as for controlling the graphene electron-gas ultrafast dynamics before relaxation and thermalization of the absorbed energy takes place.

\begin{figure}
\begin{center}
\includegraphics[width=70mm,angle=0,clip]{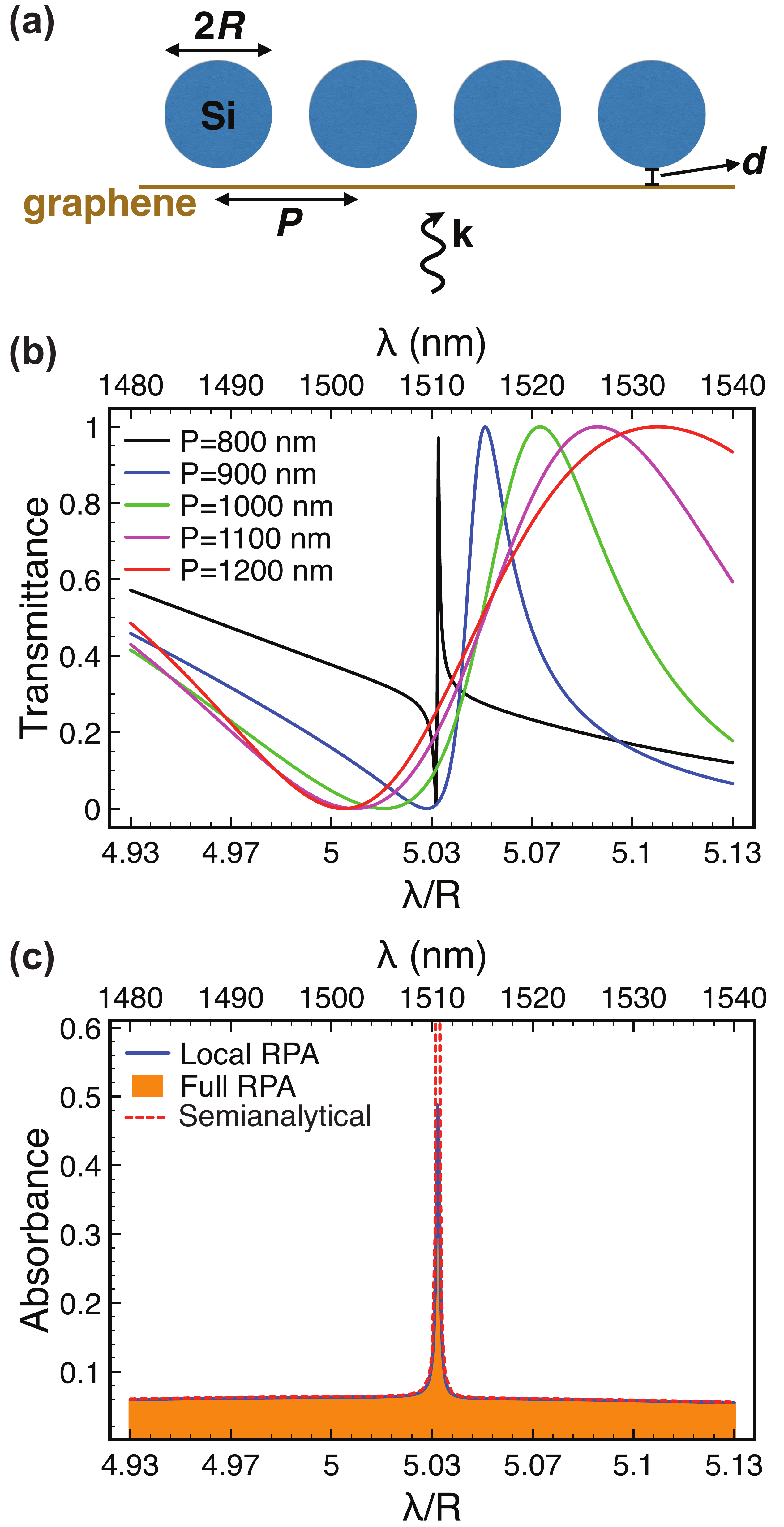}
\caption{{\bf Tunable absorption in graphene decorated with a 2D array of Mie resonators.} {\bf (a)} Geometry and parameters of a triangular array of silicon spheres near graphene. {\bf (b)} Normal-incidence transmission through the sphere array without graphene for different lattice periods $P$. The wavelength is shown both normalized to the sphere radius $R$ (lower scale) and for $R=300\,$nm (upper scale). {\bf (c)} Absorbance of the array when it is placed near undoped graphene (silicon-carbon distance $d=R/150$) under normal incidence. The lattice period is $P=800\,$nm. We compare two different approximations for the graphene conductivity (full random-phase approximation \cite{WSS06,HD07} (RPA) and
local-RPA \cite{paper235}) with a semi-analytical model, as discussed in the main text.} \label{Fig4}
\end{center}
\end{figure}

Because the maximum value of $\delta\sigma^{\rm abs}$ produced by a single silicon sphere is comparable to its projected area, we expect to obtain unity-order changes in the absorption when the graphene is decorated by a periodic array. This is illustrated in Fig.\ \ref{Fig4}, where we concentrate on the spectral region around the rightmost Mie mode of Fig.\ \ref{Fig3}a (labeled B). We consider the silicon spheres to be arranged in a triangular lattice, which we simulate using a layer-KKR approach \cite{SYM00} (see Methods). Interestingly, there is strong interaction between the particles for the lattice spacings $P$ under consideration, which can be intuitively quantified from the fact that the extinction cross section of the sphere equals the area of a circle of diameter $1.75\,\mu$m. The transmission of the particle array experiences dramatic spectral variations as $P$ is changed, eventually generating a narrow transmission peak, which is relatively close, but not on top of the lowest-order Wood anomaly, occurring when the wavelength is equal to the period at normal incidence; we thus attribute this feature to the interaction between Mie modes of the spheres, as the wavelength is close (but not right on) a lattice resonance that narrows the resulting spectral feature (see supplementary Fig.\ \ref{FigS7}). A similar mechanism leading to sharp, narrow asymmetric resonances has already been described in the context of cavity-waveguide coupling \cite{F02}. The absorbance associated with this narrow peak is boosted, approaching 50\% with undoped graphene (Fig.\ \ref{Fig4}c), whereas doped graphene shows comparatively negligible absorbance (not shown).

In the layer-KKR simulation method \cite{SYM00}, the homogeneous graphene film enters through its reflection and transmission coefficients for different diffracted orders (i.e., a collection of propagating and evanescent plane waves, each of them corresponding to a fixed value of the parallel wave vector). This allows us to use the full random-phase approximation (RPA) conductivity \cite{WSS06,HD07} $\sigma(k_\parallel,\omega)$, which includes nonlocal effects associated with finite parallel wave vectors $k_\parallel$ corresponding to those diffracted orders. Because the size of the spheres and the lattice periods under consideration are large compared with $\vF/\omega$ (i.e., the ratio of the graphene Fermi velocity to the light frequency, $\sim0.8\,$nm for a wavelength of $1.5\,\mu$m), nonlocal effects are in fact negligible, which explains why we obtain the same results on the scale of the figure by just using the value $\sigma=e^2/e\hbar$ for the conductivity in undoped graphene instead of the full RPA. The same argument explains why plasmons are not excited here in doped graphene. Additionally, we obtain very similar results with the semi-analytical model of Eq.\ (\ref{SAM}) (except very close to the resonance), which we apply by averaging the parallel electric-field intensity enhancement over a unit cell. The intensity in the semi-analytical model is calculated without the graphene, just to provide insight into the absorption process. However, when we calculate it including the carbon layer, the absorbance $A$ predicted by Eq.\ (\ref{SAM}) cannot be told apart on the scale of the figure from the one given by the far-field transmittance and reflectance (i.e., $A=1-T-R$), thus corroborating the numerical accuracy of our calculations.

For tutorial purposes, the above discussion on the coupling to Mie resonances is based on self-standing graphene, but qualitatively similar conclusions are obtained when examining graphene supported on a glass substrate (see supplementary Fig.\ \ref{FigS7}).

\begin{figure*}
\begin{center}
\includegraphics[width=120mm,angle=0,clip]{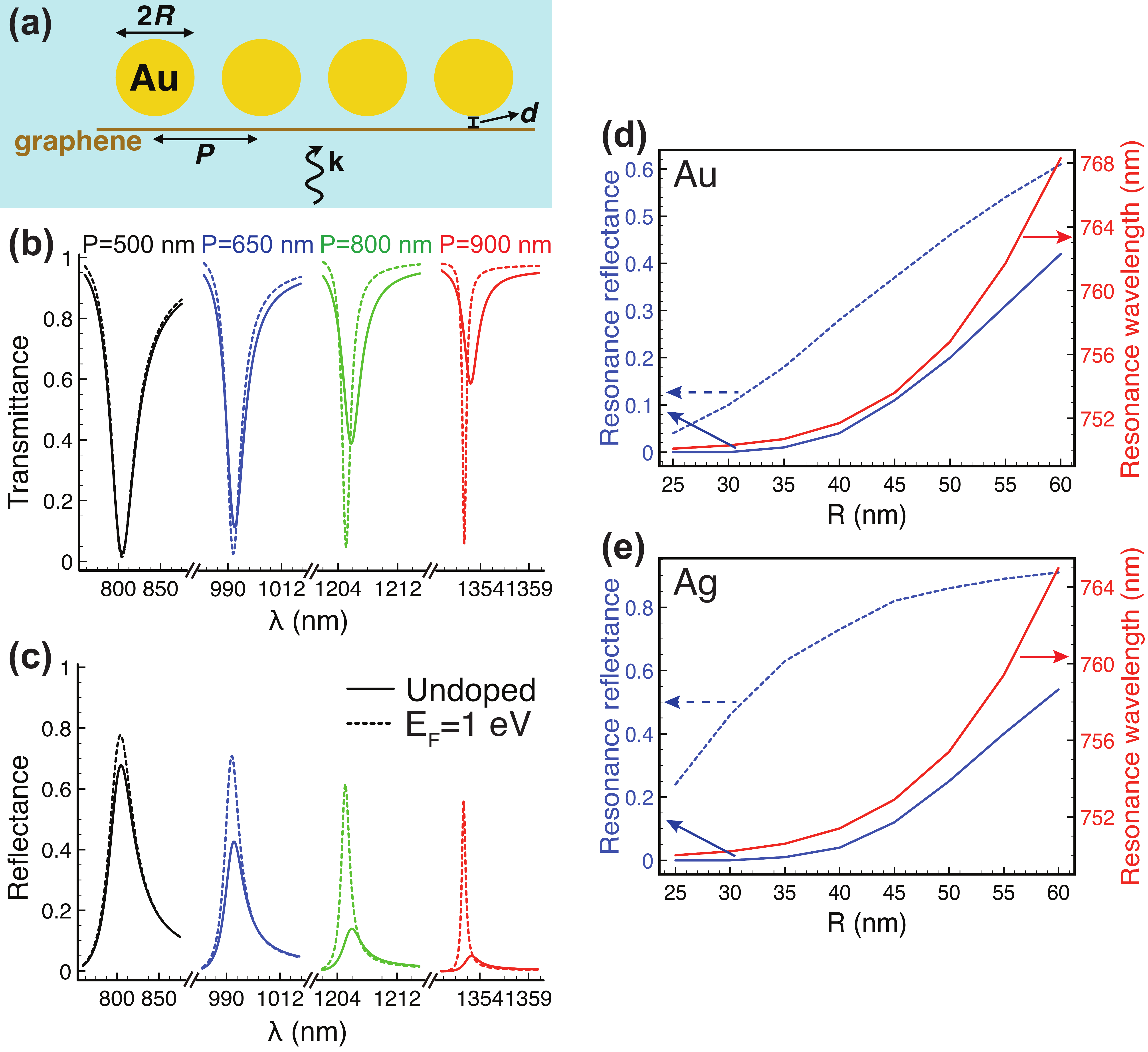}
\caption{{\bf Enhanced tunable graphene absorption by coupling to lattice resonances in 2D metal particle arrays.} {\bf (a)} Square array of gold spheres (radius $R$) placed above graphene (2\,nm gold-to-carbon separation). The entire system is assumed to be embedded in silica ($\epsilon=2.25$). {\bf (b,c)} Normal-incidence transmission (b) and reflection (c) spectra for $R=80\,$nm and different lattice periods $P$ with either doped (broken curves, $\EF=1\,$eV) or undoped (solid curves) graphene. The spectra are dominated by lattice resonances occurring near a free-space light wavelength $\lambda\sim P\sqrt{\epsilon}$. {\bf (d)} Peak wavelength with doped graphene (right scale) and transmission at that wavelength with either doped or undoped graphene (left scale) as a function of gold sphere radius for a period $P=500\,$nm. {\bf (e)} Same as (d) for silver particles.} \label{Fig5}
\end{center}
\end{figure*}

\subsection{Enhanced graphene optical absorption and switching by coupling to lattice resonances}

We now discuss the absorption enhancement produced by lattice resonances, for which strong scatterers such as metallic particles are preferable. Although metals introduce additional losses, their absorbance is relatively small in the NIR, so graphene can still make a big difference. This is corroborated in Fig.\ \ref{Fig5}, where we consider a graphene layer decorated with a 2D square array of gold spheres surrounded by silica for different values of the lattice spacing $P$. The transmission (Fig.\ \ref{Fig5}b) and reflection (Fig.\ \ref{Fig5}c) spectra of these structures exhibit sharp features emerging near the Wood anomaly condition (i.e., when the wavelength in the surrounding dielectric is close to the period, or equivalently, when a diffraction order becomes grazing), which can be easily understood in terms of lattice resonances \cite{R1907,paper090}. As the period is increased, these features move to the red, where the metal is less lossy, and consequentally, the resonances become narrower. The additional absorption produced by the undoped graphene then becomes more noticeable (see supplementary Fig.\ \ref{FigS8}), eventually causing a decrease in peak transmittance of $\sim60\%$, accompanied by a 28-fold reduction in reflectance.

\section{CONCLUSIONS AND OUTLOOK}

In conclusion, monolayer graphene can be used to produce unity-order changes in the transmission, reflection, and absorption of light down to the vis-NIR domain when comparing its electrically doped and undoped states, considering realistic levels of doping ($\EF\sim1\,$eV) that are currently attainable through electrical gating. It should be emphasized that the calculations here presented for geometries containing only graphene and dielectrics are scalable, so that the main requirement is that the graphene can be switched back and forth between $E_F=0$ and $E_F>E_{\rm p}/2$, where $E_{\rm p}$ is the photon energy under operation; provided this condition is satisfied, all geometrical lengths and the light wavelength can be scaled by a common factor, leading to the same values for the transmission and absorption. For example, the modulation at 700\,nm wavelength predicted in Fig.\ \ref{Fig1} with doping up to $E_F=1.1\,$eV can be also extrapolated to the 1550\,nm telecom wavelength with doping up to $E_F=0.5\,$eV after scaling all lengths by a factor of $\sim2$.

Interestingly, we find that undecorated graphene in a planar multilayer dielectric structure can modulate transmission near the point of resonant tunneling under total internal reflection, with absolute changes exceeding $90\%$ and an extinction ratio $>15\,$dB. Similar levels of modulation are found for graphene placed inside a realistic Fabry-Perot cavity. Large vis-NIR modulation depths are also predicted in a graphene layer decorated with periodic arrays of silicon or gold particles. Obviously, the depth of modulation is reduced by coupling to impurities in low-quality graphene, where optical losses can be still significant under high doping. Nonetheless, we find a substantial degree of modulation even in the presence of large residual absorption (e.g., $\sim50\%$ modulation in the device of Fig.\ \ref{Fig1} when the residual optical loss amounts to 14\% of the ideal absorption-free highly doped material, as shown in Fig.\ \ref{Fig1}e).

The mechanisms here considered for light modulation by graphene can be integrated in devices spanning only a few square microns in size, so they require a relatively small amount of doping charge to operate. We thus anticipate that these systems will be able to modulate vis-NIR light at high speeds with a minute consumption of power, typical of capacitive devices. This is an advantage with respect to alternative commutation devices based on quantum-wells \cite{IDA14} and phase-change materials \cite{WCL15}. For example, we envision an integrated commutation device operating over an area $A=50\times50\,\mu$m$^2$ (i.e., covering a customary optical beam size), for which we estimate a capacitance $C=A\epsilon/4\pi d\sim0.3\,$pF, where we consider $\epsilon=4$ (DC silica) and a gate separation $d=300\,$nm (notice that there is great flexibility in the choice of $d$ in some of our devices). The time response is then limited by the sheet resistance of the graphene layer ($\sim100\,\Omega/\Box$), giving an overall cutoff frequency for the electrical bandwidth of $1/2\pi RC\sim5\,$GHz, while the optical limit for the electrical modulation of the photonic response (i.e., the effect related to the decay time of the resonance) renders a larger cutoff ($c/2LQ\sim150\,$GHz for a cavity length $L\sim1\,\mu$m and a quality factor $Q\sim10^3$). The large electro-optical response of graphene combined with its small volume are thus ideal attributes for the design of fast optical modulators and switches operating in the vis-NIR, which can benefit from the coupling to optical resonators such as those explored in the present work. In particular, the planar structures presented in Figs.\ \ref{Fig1} and \ref{Fig2}, which rely on unstructured graphene, provide relatively affordable designs that are appealing for micro integration and mass production.

\section*{METHODS}

{\bf Schr\"odinger equation for TE polarized light in a planar multilayer structure.} For our purpose, it is convenient to write the wave equation for the electric field $\Eb$ as $\nabla\times\nabla\times\Eb-k^2\epsilon\,\Eb=0$, where $k=\omega/c$ is the free-space light wave vector and $\epsilon$ is the frequency and position dependent local dielectric function. We consider a structure formed by several planar layers (see Fig.\ \ref{Fig1}), illuminated under TE polarization (i.e., with the electric field parallel to the planes), and oriented with the $z$ axis perpendicular to the interfaces, so that the spatial dependence of the dielectric function is only through $z$ (i.e., $\epsilon(z)$). Then, we can write the electric field as $\Eb(\rb)=\psi(z)e^{i k_\parallel y}\xx$, where $k_\parallel=k\sqrt{\epsilon_{\rm i}}\sin\theta_{\rm i}$ is the parallel wave vector component, which is determined by the angle of incidence $\theta_{\rm i}$ at the near medium of permittivity $\epsilon_{\rm i}$. With this form of the electric field, the wave equation reduces to
\begin{equation}
\frac{-1}{2}\frac{d^2\psi(z)}{dz^2}+V(z)\psi(z)=E\psi(z), \nonumber
\end{equation}
where we have defined the equivalent potential $V(z)=[1-\epsilon(z)]k^2/2$ and the effective energy $E=(k^2-k_\parallel^2)/2$. Interestingly, metals ($\epsilon<0$) and dielectrics ($\epsilon>1$) produce repulsive ($V>0$) and attractive ($V<0$) potentials, respectively, in this equivalent Schr\"odinger model.

{\bf Graphene conductivity.} We model graphene through its 2D conductivity. For doped graphene, we use a previously reported local-RPA model \cite{FV07,FP07_2}, conveniently corrected to account for finite temperature $T$ in both intra- and interband transitions \cite{paper235}. More precisely,
\begin{align}
\sigma(\omega)=&\frac{e^2}{\pi\hbar^2}\frac{\ii}{(\omega+\ii\tau^{-1})}
\nonumber\\
&\times\left[\EFT-\int_0^\infty dE\;\frac{f_E-f_{-E}}{1-4E^2/[\hbar^2(\omega+\ii\tau^{-1})^2]}\right],
\nonumber
\end{align}
where $\EF$ is the Fermi energy, \[\EFT=\EF+2k_{\rm B}T\,\log\left(1+\ee^{-\EF/k_{\rm B}T}\right)\] effectively accounts for thermal corrections in the doping level, and $f_E=1/[1+\ee^{(E-\EF)/k_{\rm B}T}]$ is the Fermi-Dirac distribution function. For undoped graphene, the above expression converges to the well-known limit $\sigma=e^2/4\hbar$. We further account for finite wave-vector effects (nonlocality) through the full-RPA model \cite{WSS06,HD07}, in which $\sigma$ depends on $k_\parallel$ and $\omega$, but we find those effects to be negligible (see Fig.\ \ref{Fig4}), as expected from the large mismatch between the Landau damping range $\vF/\omega$ at vis-NIR frequencies $\omega$ and both the light wavelength and the distances involved in the structures under consideration. Throughout this work, we take $T=300\,$K and assume an inelastic decay time given by the Drude model \cite{AM1976,JBS09} (i.e., $\tau=\mu\EF/ev_{\rm F}^2$) with an impurity limited DC mobility $\mu=2000\,$cm$^2/($V\,s$)$.

{\bf Multilayer structure simulation.} The transmission and field enhancement of planar multilayers are obtained through a standard transfer matrix approach. In particular, we use the reflection and transmission coefficients for a plane wave of parallel wave vector $k_\parallel$ incident from medium 1 on a graphene layer of conductivity $\sigma$ placed at the interface with another medium 2, which upon direct solution of Maxwell's equations for the $s$ (TE) polarization under consideration are found to be \cite{J99} $r^s_{12}=(k_{z1}-k_{z2}+g_s)/(k_{z1}+k_{z2}+g_s)$ and $t^s_{12}=2k_{z1}/(k_{z1}+k_{z2}+g_s)$, respectively, where $g_s=4\pi\sigma\omega/c^2$ and $k_{zj}=\sqrt{k^2\epsilon_j-k^2_\parallel+\ii0^+}$. For completeness, we give the coefficients for $p$ (TM) polarization: $r^p_{12}=(\epsilon_2k_{z1}-\epsilon_1k_{z2}+g_p)/(\epsilon_2k_{z1}+\epsilon_1k_{z2}+g_p)$ and $t^p_{12}=2\sqrt{\epsilon_1\epsilon_2}k_{z1}/(\epsilon_2k_{z1}+\epsilon_1k_{z2}+g_p)$, where $g_p=4\pi\sigma k_{z1}k_{z2}/\omega$. Incidentally, the sign of the square root is chosen to yield positive real values. These expressions also describe the coefficients of interfaces without graphene, simply by taking $g_s=g_p=0$.

For tunneling transmission (Fig.\ \ref{Fig1}), the resonant wavelength of maximum transmission $\lambda_{\rm res}$ is only slightly changed from the central waveguide Fabry-Perot resonance condition, $k_{z2}d+\varphi=N\pi$, where $d$ is the waveguide thickness, $N$ ($=0$ under the conditions of this work) is the order of the resonance, and $\varphi=\arg\{r_{21}^s\}$ (we choose media 1 and 2 right outside and inside the waveguide, respectively). To linear order in $\sigma$, we find \begin{equation}
\lambda_{\rm res}=\frac{-2\pi k_{z2}d}{k\varphi^2}\,\left(\varphi+\frac{8\pi (k/c)\cos(\varphi/2){\rm Im}\{\sigma\}}{\sqrt{|k_{z1}|^2+|k_{z2}|^2}}\right).
\label{lambdares}
\end{equation}
Incidentally, $\sigma$ has units of velocity in CGS, so this expression is dimensionally correct.

{\bf Particle arrays.} We use the layer-KKR method to simulate periodic particle arrays near planar interfaces \cite{SYM00}. This method relies on an expansion of the electromagnetic field in terms of spherical vector waves around the particles and plane waves near the graphene. The scattering by the spheres then involves multiplication by Mie coefficients, whereas the graphene enters through its reflection coefficients (see above). Plane and spherical waves are analytically transformed into each other, giving rise to a self-consistent system of equations projected on the coefficients of the sphere multipoles. Translational lattice symmetry is used to reduce the number of plane waves to those of a discrete set corresponding to different diffraction orders (i.e., two waves of orthogonal polarizations for each reciprocal lattice vector). We achieve convergence with $\sim100$ such waves and neglecting sphere multipoles of orbital angular momentum number above 7. This method directly yields the reflection, transmission, and absorption coefficients used to produce Figs.\ \ref{Fig4} and \ref{Fig5} for periodic particle arrays near planar interfaces including graphene.

\section*{Acknowledgement}

This work has been supported in part by the European Commission (contract Nos. Graphene Flagship CNECT-ICT-604391 and FP7-ICT-2013-613024-GRASP) and the "Fondo Europeo de Desarrollo Regional" (FEDER, contract No. TEC2013-46168-R).


\newpage
\widetext

\begin{figure}
\begin{center}
\includegraphics[width=0.6\textwidth]{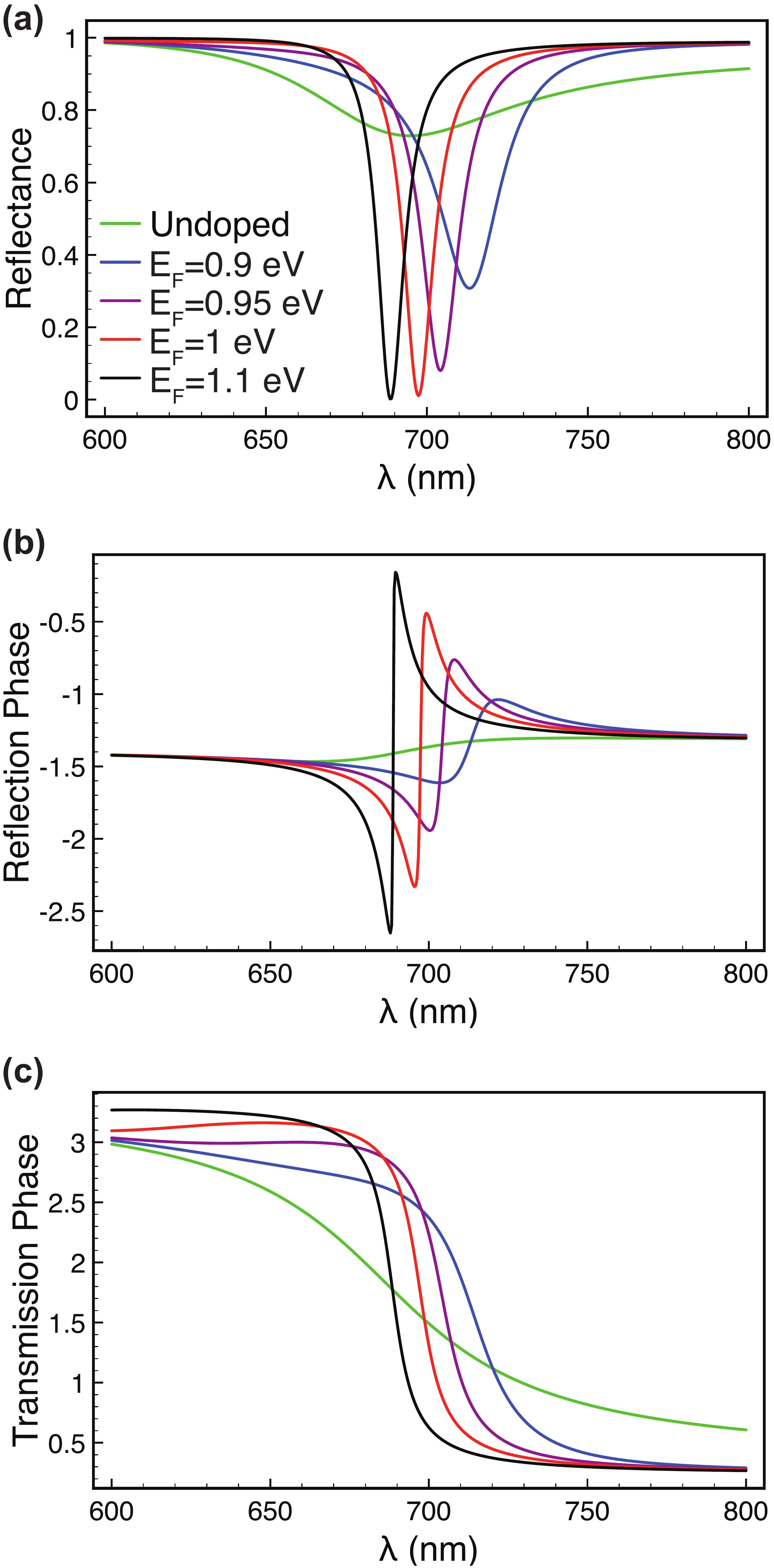}
\caption{({\it Supplementary figure.}) Reflectance (a), reflection phase (b), and transmission phase (c) under the same conditions as in Fig.\ 1e of the main paper.} \label{FigS1}
\end{center}
\end{figure}

\begin{figure}
\begin{center}
\includegraphics[width=0.6\textwidth]{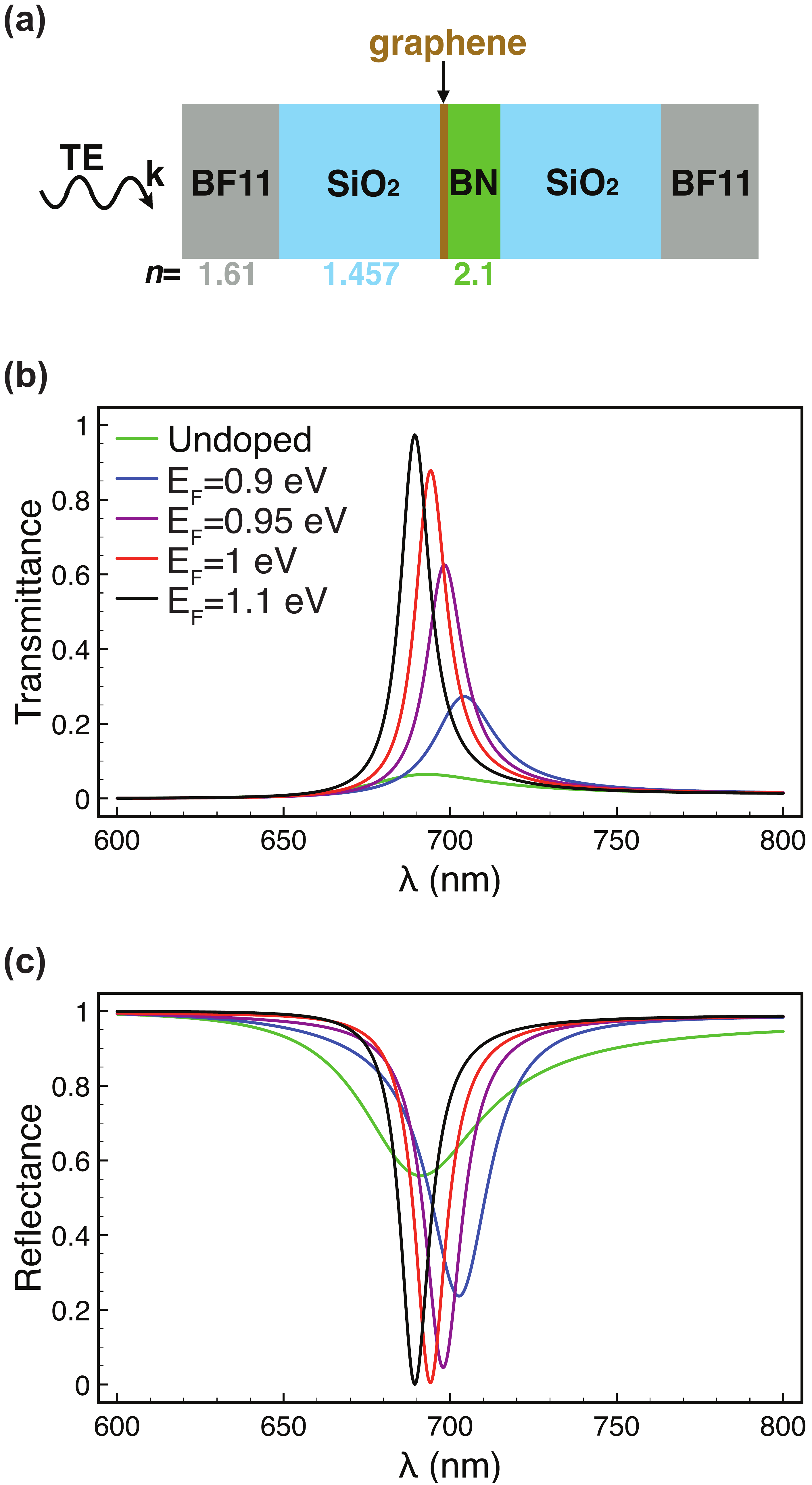}
\caption{({\it Supplementary figure.}) Transmittance (b) and reflectance (c) for a structure similar to that of Fig.\ 1 of the main paper, but containing only one graphene layer (see (a)), under the same conditions of light incidence.} \label{FigS2}
\end{center}
\end{figure}

\begin{figure}
\begin{center}
\includegraphics[width=0.6\textwidth]{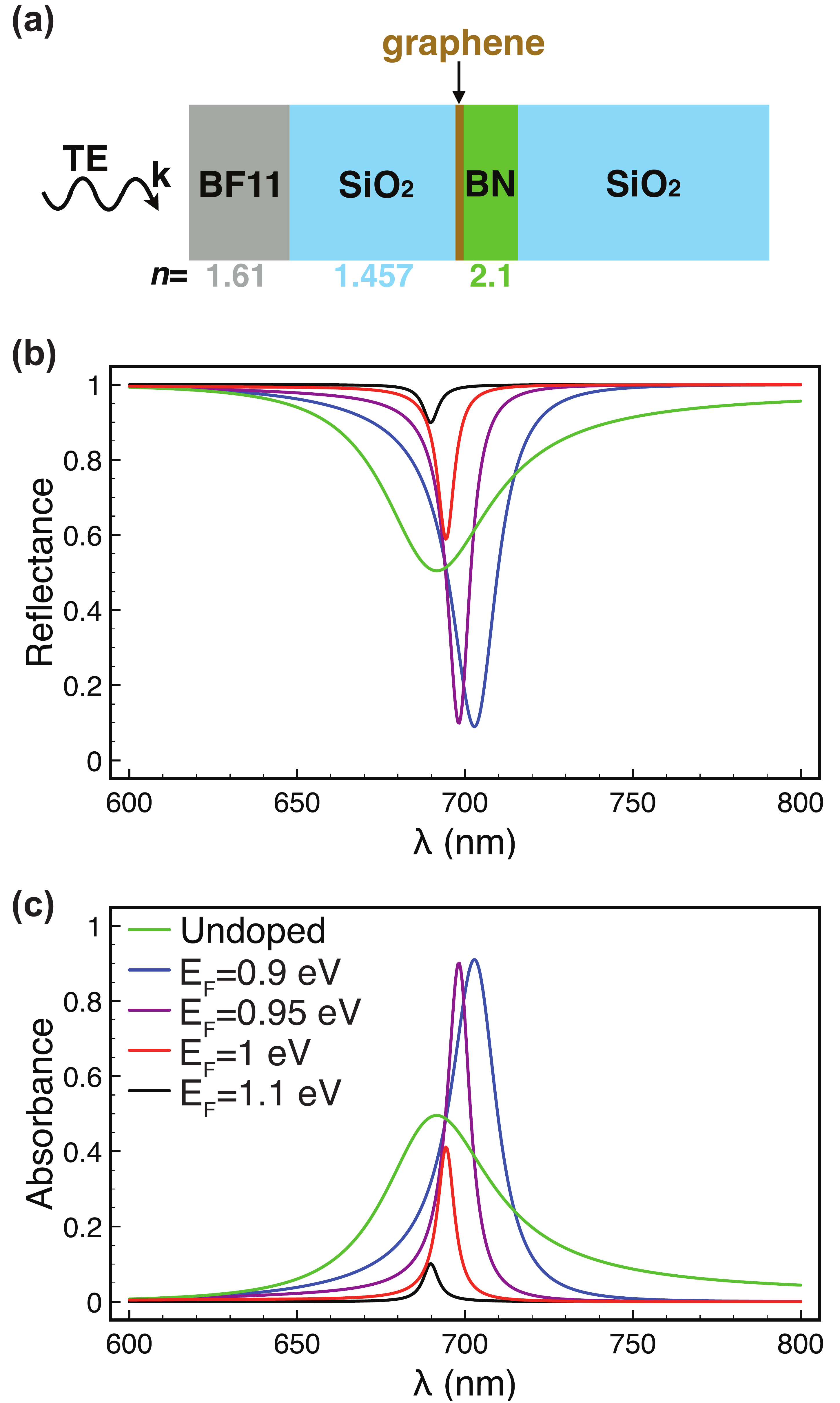}
\caption{({\it Supplementary figure.}) Reflectance (b) and absorbance (c) of a similar structure as in Fig.\ 1 of the main paper, but containing only one graphene layer and without any out-coupling BF11 medium on the right-hand side of the structure. The conditions of light incidence are the same as in Fig.\ 1.} \label{FigS3}
\end{center}
\end{figure}

\begin{figure}
\begin{center}
\includegraphics[width=0.6\textwidth]{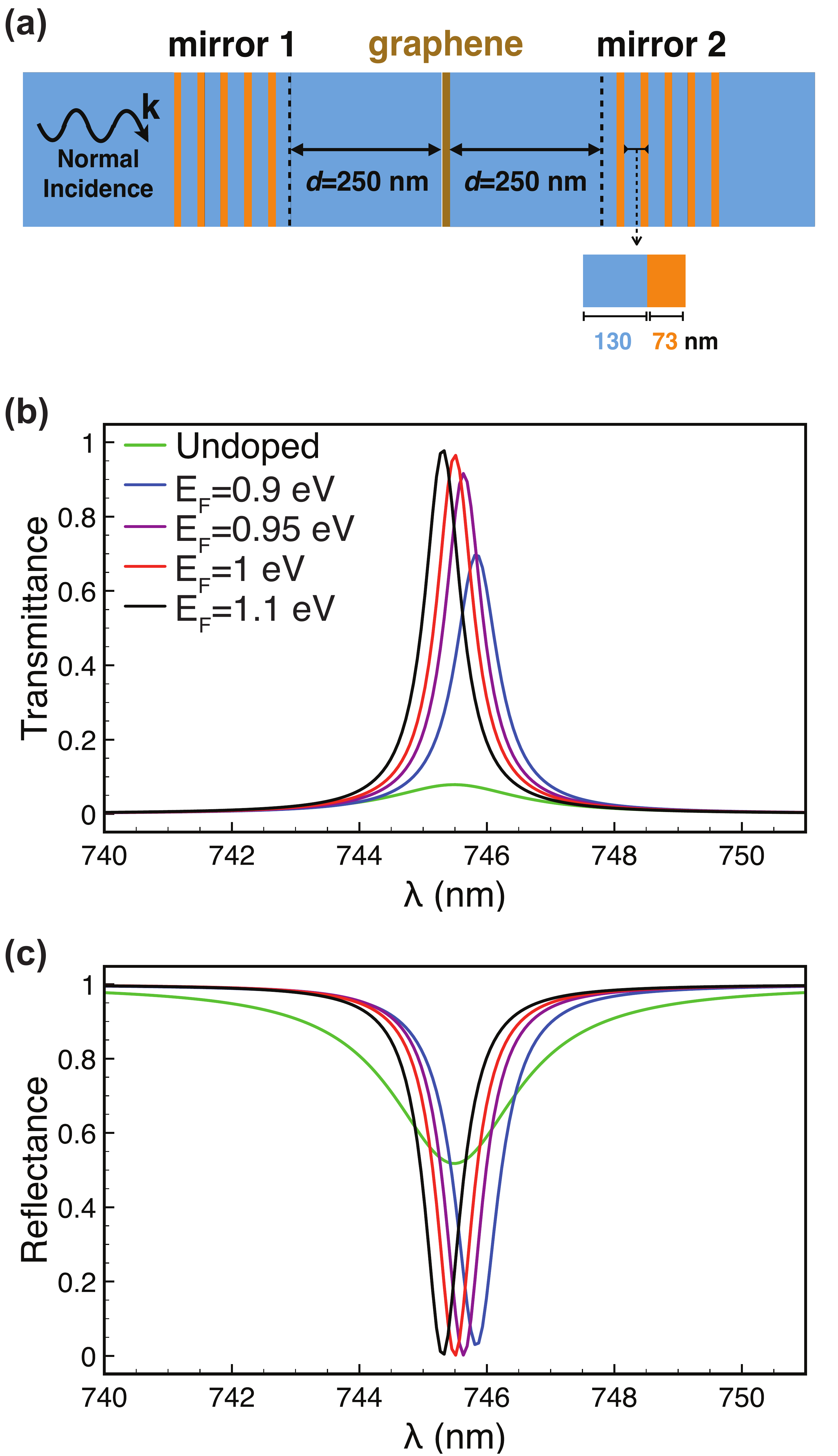}
\caption{({\it Supplementary figure.}) Performance of a Fabry-Perot cavity similar to that of Fig.\ 2 of the main paper, but filled with glass and designed to operate in the same spectral region using modified geometrical parameters.} \label{FigS4}
\end{center}
\end{figure}

\begin{figure}
\begin{center}
\includegraphics[width=0.7\textwidth]{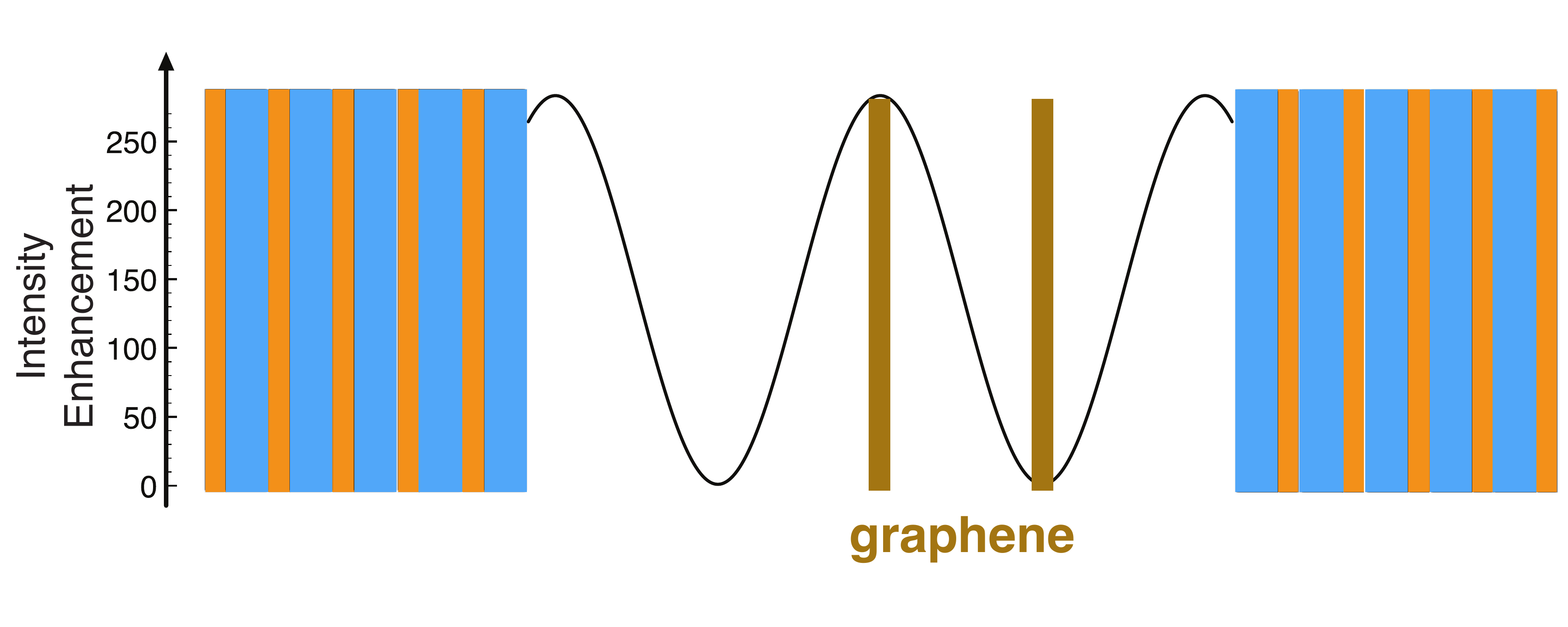}
\caption{({\it Supplementary figure.}) Electric field intensity enhancement relative to the incident intensity inside the Fabry-Perot cavity considered in Fig.\ 2 of the main paper, calculated at the 738\,nm resonance wavelength in the absence of graphene. The addition of a second graphene layer at an antinode (rightmost graphene layer in this plot) produces exactly the same transmission and reflection spectra as in Fig.\ 2, regardless the doping state of the extra layer. The width of the cavity is 800\,nm and other geometrical parameters are the same as in Fig.\ 2a.} \label{FigS5}
\end{center}
\end{figure}

\begin{figure}
\begin{center}
\includegraphics[width=0.7\textwidth]{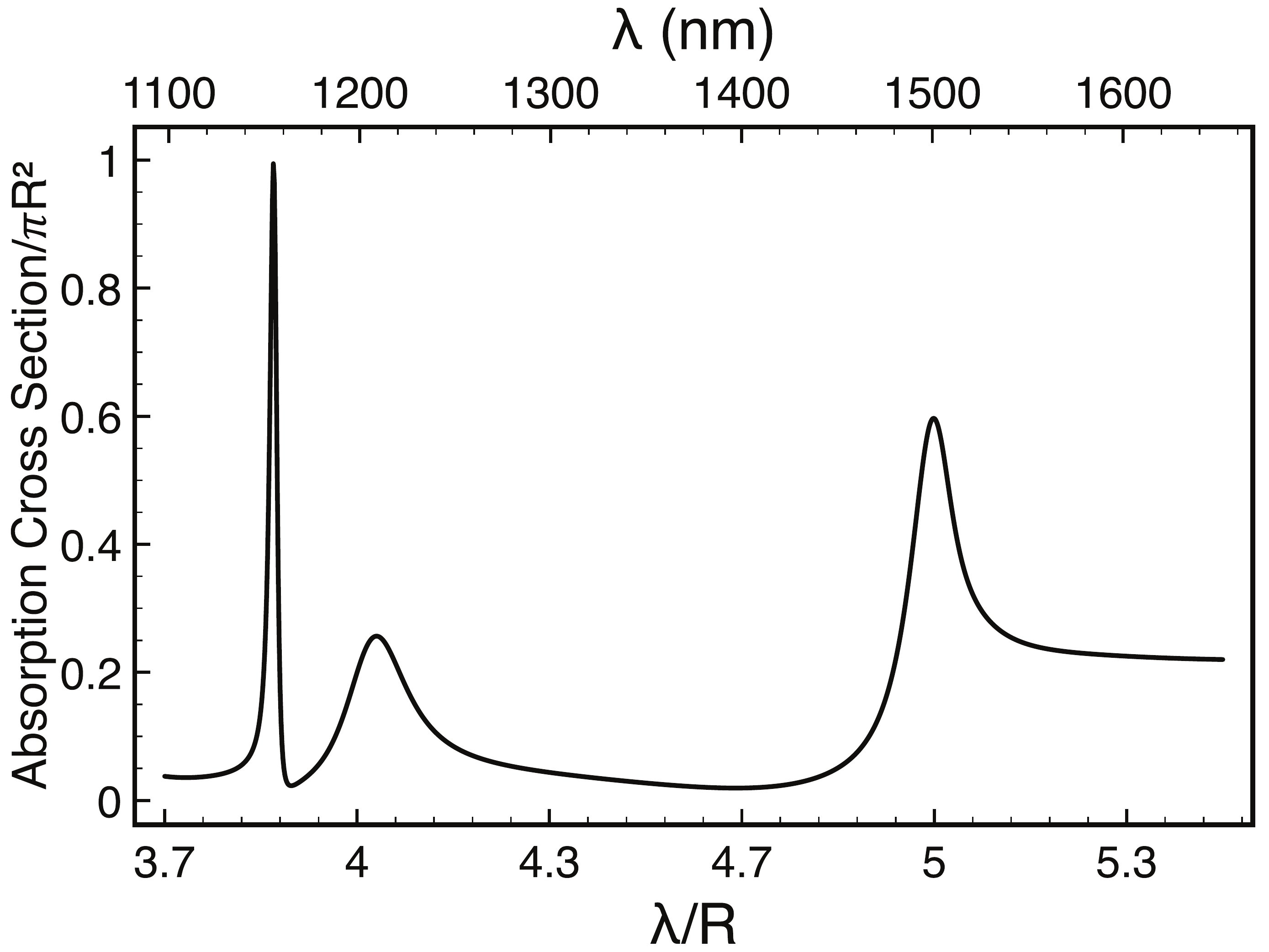}
\caption{({\it Supplementary figure.}) Graphene-doping-induced change is the dark-field scattering cross section (i.e., integrated over scattering directions other than the specular reflection or direct forward transmission) of a silicon sphere under the same conditions as in Fig.\ 3a. The curve represents the difference in angle-integrated elastic cross section when the graphene is undoped or doped to $E_{\rm F}=1\,$eV, calculated using a modal expansion described elsewhere \cite{paper200}.}\label{FigS6}
\end{center}
\end{figure}

\begin{figure}
\begin{center}
\includegraphics[width=0.5\textwidth]{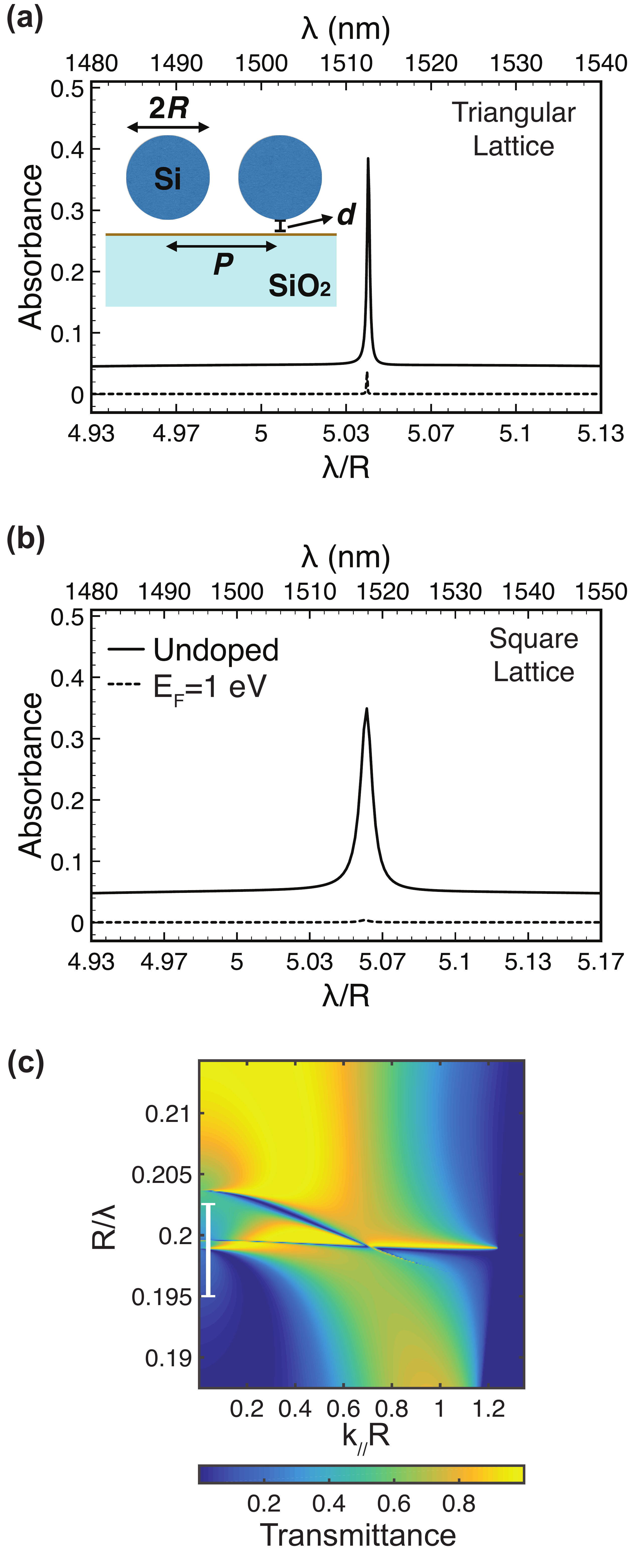}
\caption{({\it Supplementary figure.}) {\bf (a)} Normal-incidence ($k_\parallel=0$) absorption spectra for a triangular lattice of silicon spheres (radius $R=300\,\mathrm{nm}$ and lattice period $P=800\,\mathrm{nm}$) placed on top of a graphene sheet (silicon-carbon separation distance $d=2\,\mathrm{nm}$) when the graphene is supported on a silica substrate. {\bf (b)} Same as (a) for a square lattice. {\bf (c)} Dispersion diagram of the triangular silicon-sphere lattice without graphene in the Mie resonance region under consideration. The white vertical segment in (c) indicates the spectral range in (a), dominated by a sphere Mie mode that is crossed by a lattice resonances at finite $k_\parallel$. The lattice resonance produces a narrowing of the Mie mode.} \label{FigS7}
\end{center}
\end{figure}

\begin{figure}
\begin{center}
\includegraphics[width=0.7\textwidth]{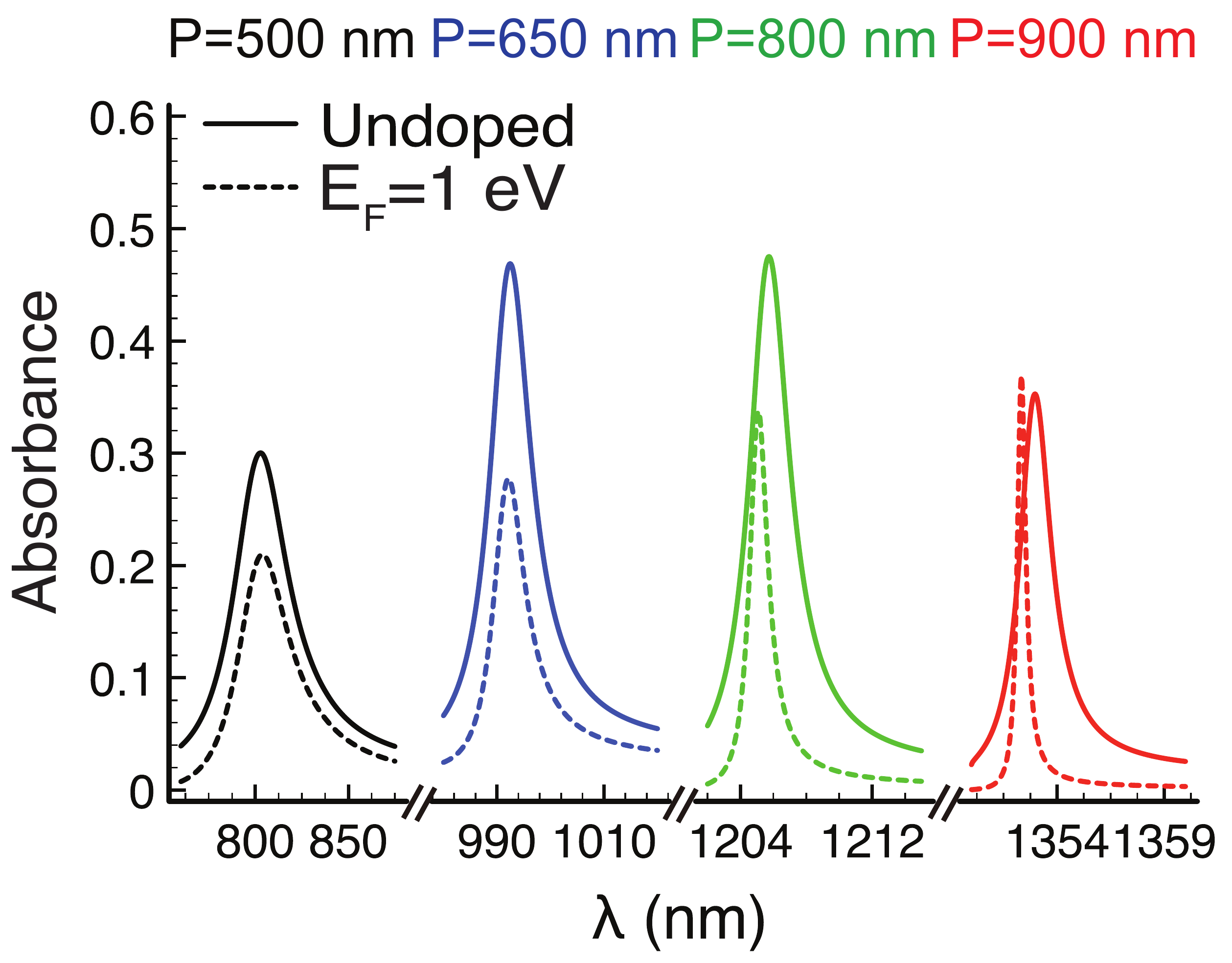}
\caption{Absorbance spectra under the same conditions as in Fig.\ 5b,c of the main paper.} \label{FigS8}
\end{center}
\end{figure}

\end{document}